\documentclass[10pt,english, aps,superscriptaddress,showpacs,floatfix,prl,lengthcheck,floatfix,nofootinbib,twocolumn]{revtex4-2}

\usepackage[T1]{fontenc}
\usepackage[utf8]{inputenc}
\usepackage{amsmath}
\usepackage{amssymb}
\usepackage{graphicx}
\PassOptionsToPackage{normalem}{ulem}
\usepackage{ulem}
\usepackage{graphicx} 
\usepackage{float}
\usepackage[utf8]{inputenc}
\makeatletter

\usepackage[T1]{fontenc}
\usepackage[utf8]{inputenc} 

\usepackage{braket}

\usepackage{amsmath}

\usepackage{color}
\usepackage{babel}
\usepackage{verbatim}
\usepackage{amsmath}
\usepackage{amssymb}
\usepackage{graphicx}
\usepackage[pdfusetitle,colorlinks,linkcolor=blue,citecolor=blue,urlcolor=blue]{hyperref}
\usepackage{microtype}

\makeatother

\usepackage{babel}
\begin{document}
\title{Valley polarization driven by two-color circular fields from a strong-field tunneling perspective}
\author{Rui E.F. Silva}
\affiliation{\emph{Instituto de Ciencia de Materiales de Madrid (ICMM), Consejo
Superior de Investigaciones Científicas (CSIC), Sor Juana Inés de
la Cruz 3, 28049 Madrid, Spain}}
\author{Olga Smirnova}
\affiliation{\emph{Max Born Institute, Max-Born-Straße 2A, 12489, Berlin, Germany}}
\author{Misha Ivanov}
\affiliation{\emph{Max Born Institute, Max-Born-Straße 2A, 12489, Berlin, Germany}}
\author{\'Alvaro Jim\'enez-Gal\'an}
\email{alvaro.jimenez@csic.es}
\affiliation{\emph{Instituto de Ciencia de Materiales de Madrid (ICMM), Consejo
Superior de Investigaciones Científicas (CSIC), Sor Juana Inés de
la Cruz 3, 28049 Madrid, Spain}}

\begin{abstract}

Valley polarization in hexagonal materials driven by the form of the lightwave and its orientation relative to the lattice, rather than solely by the helicity of the driver, has recently been demonstrated both theoretically and experimentally. This has extended valley control to the non-resonant, sub-cycle regime and to inversion-symmetric materials. One interpretation of this field-orientation-dependent valley polarization is that the laser-dressed, cycle-averaged band structure obtained from a Floquet-type approach is modified so as to lift the degeneracy between the valleys. Here, we derive a complementary explanation based on strong-field tunneling dynamics. We show that the effective valley gap at the dominant injection times depends on the orientation of the field relative to the lattice through the trigonal-warping term of the low-energy dispersion. We derive general expressions for co- and counter-rotating $\omega+N\omega$ fields and show that the leading orientation-dependent contribution survives the cycle average only for field configurations compatible with the threefold lattice symmetry, most notably counter-rotating $\omega+2\omega$ and co-rotating $\omega+4\omega$ fields.

\end{abstract}
\maketitle

\section{Introduction}
Hexagonal 2D materials with broken inversion symmetry such as transition metal dichalcogenides or hexagonal boron nitride, host two energy-degenerate valleys, $K$ and $K'$, that couple differently to circularly-polarized light~\cite{vitale2018, schaibley2016valleytronics, mak2012control}. These valley optical selection rules arise due to the $C_3$ symmetry of the lattice, and form the basis of valleytronics. Ever since its theoretical prediction by Xiao et al.~\cite{xiao2007valley, yao2008valley, xiao2010berry}, many works have now established valley selection rules~\cite{mak2012control,zeng2012valley}, and used valley physics as a way to explore other phenomena such as exciton decoherence times, coherent control of valley pseudospin or valley filters~\cite{hao2016direct, Gucci2026, ye2017optical,ang2017valleytronics,rana2023all, PhysRevB.111.155421}.

Recently, it was shown that valley-dependent physics is not restricted to the use of resonant circularly-polarized pulses. Under intense, low-frequency driving, linearly-polarized single-cycle pulses and symmetry-tailored pulses have been shown to produce strong valley dichroism and photocurrents in materials with and without inversion symmetry~\cite{jimenez2021sub, jimenez2020lightwave,mrudul2021controlling, lesko2026probing, mrudul2021light, PhysRevB.111.155421}. In particular, the bicircular field - a combination of a circularly-polarized fundamental $\omega$ field and its counter-rotating second harmonic $2\omega$ -  has been shown to switch valley polarization in monolayer hexagonal boron nitride and bulk MoS$_2$ through spatial rotation of its light form~\cite{mitra2024light,tyulnev2024valleytronics}.

One way to understand this breaking of valley symmetry of the driven system is through the laser-induced dynamical symmetries of the time-periodic system~\cite{oka2009photovoltaic,gomez2013floquet,cayssol2013floquet,rudner2020band,neufeld2019floquet, batista2026field}. Within an adiabatic Floquet approach, such fields have been shown to induce a complex second neighbour hopping that lifts the degeneracy between the valleys, thus inducing valley polarization~\cite{jimenez2020lightwave,molinero2024manipulation}, similar to the case of the Haldane model~\cite{haldane1988model}. This lightwave-induced valley polarization is dominated by the orientation of the field's vector potential relative to the lattice. This is in contrast to other mechanisms of inducing laser-dressed material parameter modifications, such as those in graphene~\cite{oka2009photovoltaic,mciver2020light}, where it is dominated by the helicity of the driving field.

In the strong field regime, however, ionization dynamics are more faithfully modeled using a tunneling approach~\cite{Keldysh:1965ojf}. Here, we show how valley bandgap asymmetry and polarization arises also in a quasistatic tunneling framework, which is particularly suited for strong and few-cycle pulses such as those used in high-harmonic generation and lightwave electronics~\cite{heide2024petahertz,ghimire2019high,Ciappina_2017,PhysRevLett.113.073901}. We demonstrate that valley polarization due to the orientation of the lightwave appears once the trigonal term is considered in the low-energy expansion of the band structure close to the valleys. We derive an analytical expression of the laser-dressed valley bandgaps at the optimal tunneling times for co- and counter-rotating combinations of $\omega+N\omega$ fields. For the bicircular field, made of counter-rotating $\omega+2\omega$ fields, we further show in Appendix~\ref{sec:helicity} that in the weak-field limit, a transformation to the rotating frame of reference translates the helicity-dependent valley selection rule into a Coriolis mass term in the valley gap energy, similar to the result obtained in atomic photoionization from $p$-orbitals~\cite{pisanty2017strong}.

Our derivation shows that only particular combination of colors generates such orientation-dependent valley dichrosim when averaged over a cycle, and how one can control the valley polarization with the relative strength between the two colors. The derivation is applicable to bilayer and bulk systems in the 2H phase, and also agrees with previous work in graphene~\cite{mrudul2021light}. We support our analytical derivation with time-dependent simulations of strong-field driven dynamics in gapped graphene, demonstrating some of the predictions of the model.

\section{Valley-dependent injection from a quasistatic tunneling model}
In the strong-field tunneling regime, electron injection contributing to the final population at crystal momentum $\mathbf{k}$ is concentrated near the minima of the tunneling exponent $S(\mathbf{k},t)$ for the trajectory $\mathbf{k}(t) = \mathbf{k}+\mathbf{A}(t)$, which ends at $\mathbf{k}$ when the vector potential of the time-dependent field $\mathbf{A}(t\to \infty)=0$. We estimate the dominant electron-injection times using a tunneling exponent based on the quasistatic Keldysh expression for solids~\cite{Keldysh:1965ojf}, approximating the effective gap by the instantaneous band separation along the accelerated crystal-momentum trajectory,
\begin{equation}\label{eq:tunnel_prob}
\begin{split}
    &W(\mathbf{k},t) \propto e^{-S(\mathbf{k},t)}, \\  
    &S(\mathbf{k},t) = \frac{\pi}{2} \frac{\sqrt{\mu(\mathbf{k},t)}\Delta^{3/2}(\mathbf{k},t)}{|\mathbf{E}(t)|}
\end{split}
\end{equation}
Above, $\Delta(\mathbf{k},t)$ is the instantaneous band separation sampled by the trajectory, $\mu_\tau(\mathbf{k},t)$ is the electron-hole reduced mass along the tunneling direction and $|\mathbf{E}(t)|$ is the magnitude of the electric field. The dominant injection times are those that maximize the tunneling probability and thus minimize the tunneling exponent. In the following, we seek to find those times in a gapped graphene system driven by a two-color circular field and compare the maximum tunneling probability at each of the valleys.

For this, it is sufficient and more transparent to consider only up to the lowest-order anisotropic contribution to the band separation that distinguishes the two valleys, i.e., the trigonal warping term~\cite{xiao2007valley,castro2009electronic}. Up to this contribution, the instantaneous band separation of the trajectory that ends at the valley center is (see Appendix~\ref{sec:low-energy} for details),
\begin{equation}\label{eq:band_separation_warping}
\begin{split}
    &\Delta_\tau(t)\\
    &=2 \sqrt{m^2+v_F^2\,|\mathbf{A}(t)|^2 - \frac{v_F^2\,a}{2} \tau |\mathbf{A}(t)|^3 \sin 3\theta_{A(t)} + O^4},
\end{split}
\end{equation}
where $\pm m$ is the on-site energy on sites A and B, respectively, $v_F=\frac{3ta}{2}$ is the Fermi velocity with $t$ the real first-neighbour hopping and $a$ the nearest-neighbour distance. The valley index $\tau=\pm 1$ identifies the $K$ and $K'$ valley centers, respectively. The angle
\begin{equation}
    \theta_{A(t)} = \arg \{A_x(t) + i A_y(t)\}.
\end{equation}

In order to obtain a simple closed expression for the minimum of the valley-dependent tunneling exponent for the trajectory that ends in the valley center, we take the following approach. First, we determine the times that produce a minima of the isotropic tunneling exponent,
\begin{equation}\label{eq:isotropic_tunnel_exp}
\begin{split}
    &S(t) \approx S_0(t) = \frac{\pi}{2} \frac{\sqrt{\mu_0}\Delta_0^{3/2}(t)}{|\mathbf{E}(t)|},
\end{split}
\end{equation}
which is computed by using the first-order isotropic expression for the band separation (see Appendix~\ref{sec:low-energy}), 
\begin{equation}
\Delta_0 (t) = 2 \sqrt{m^2 + v_F^2 |\mathbf{A}(t)|^2},
\end{equation}
and the fixed electron-hole reduced mass $\mu_0$ of the isotropic Hamiltonian (see Eq.~\ref{eq:eh_mass} of Appendix~\ref{sec:low-energy}). These times, which we denote hereafter as $t_n^\text{min}$, are then used to calculate the valley-dependent tunneling exponent containing the trigonal warping contribution,
\begin{equation}\label{eq:tunnel_exponent_warping}
    S_\tau(t_n^{\text{min}}) = \frac{\pi}{2}\frac{\sqrt{\mu_0}\,\Delta_\tau^{3/2}(t_n^{\text{min}})}{|\mathbf{E}(t_n^{\text{min}})|},
\end{equation}
with $\Delta_\tau(t)$ given in Eq.~\ref{eq:band_separation_warping}. This approach gives the minima of the tunneling exponent up to first order in the trigonal warping contribution assuming fixed reduced electron-hole mass (see Appendix~\ref{sec:warping_correction}).

We thus require for the isotropic term,
\begin{equation}\label{eq:saddles}
\frac{d S_0(t)}{dt}\Bigg|_{t=t_n^{\text{min}}}=0, \quad    \frac{d^2S_0(t)}{dt^2}\Bigg|_{t=t_n^{\text{min}}}>0.
\end{equation}
The former reads
\begin{equation}\label{eq:saddle_eqs}
    C(t_n^{\text{min}}) \frac{d {|\mathbf{A}}(t)|^2}{d t}\Bigg|_{t=t_n^{\text{min}}} = \frac{d|{\mathbf{E}}(t)|^2}{dt}\Bigg|_{t=t_n^{\text{min}}},
\end{equation}
with
\begin{equation}
    C(t_n^{\text{min}}) = \frac{3v_F^2 |\mathbf{E}(t_n^{\text{min}})|^2}{2 \left[m^2+v_F^2 |\mathbf{A}(t_n^{\text{min}})|^2 \right]}.
\end{equation}

We now consider driving by a bicircular field defined as~\cite{pisanty2017strong}
\begin{equation}\label{eq:bicircular_field}
    \mathbf{E}(t) = \text{Re}\left( |E_\omega|\,e^{- i \chi \omega t}\,\hat{\mathbf{e}}_+ + |E_{2\omega}|e^{i\phi}\,e^{-2i\chi\omega t}\,\hat{\mathbf{e}}_-\right),
\end{equation}
where the circular unit vectors are defined as $\hat{\mathbf{e}}_\pm = \mp (\hat{\mathbf{e}}_x \pm i\mathbf{\hat{e}}_y)/\sqrt{2}$, and the phase $\phi$ is the two-color phase delay that also controls the orientation of the field relative to the lattice.
The parameter $\chi=\pm 1$ fixes the helicity of the fundamental component and controls the direction in which the Lissajous curve is traversed as a function of time. In the weak-field resonant limit, valley populations are governed by this helicity~\cite{xiao2007valley}; we show this for the bicircular field in Appendix~\ref{sec:helicity} using a rotating-frame approach similar to that in reference~\cite{pisanty2017strong}.
The corresponding vector potential is defined as
\begin{equation}
   \mathbf{E}(t) = -\partial_t \mathbf{A}(t),
\end{equation}
so
\begin{equation}\label{eq:bicircular_vpot}
    \mathbf{A}(t) = \text{Re}\left[ \frac{|E_\omega|}{i\chi\omega}e^{-i\chi \omega t}\,\hat{\mathbf{e}}_+  + \frac{|E_{2\omega}|e^{i\phi}}{i2\chi\omega}e^{-2i\chi \omega t}\,\hat{\mathbf{e}}_-\right].
\end{equation}
The squared values of both electric field and vector potential is
\begin{equation}\label{eq:fields_squared}
    \begin{split}
        &|\mathbf{E}(t)|^2 = \frac{|E_\omega|^2+|E_{2\omega}|^2}{2} - |E_\omega||E_{2\omega}|\cos(3\chi\omega t - \phi),\\
        &|\mathbf{A}(t)|^2 = \frac{|E_\omega|^2+|E_{2\omega}|^2/4}{2\omega^2} + \frac{|E_\omega||E_{2\omega}|}{2\omega^2}\cos(3\chi\omega t - \phi),
    \end{split}
\end{equation}
so that
\begin{equation}
\begin{split}
    \frac{d|\mathbf{E}(t)|^2}{dt} &= - 2\omega^2\,\frac{d |\mathbf{A}(t)|^2}{dt} \\
    &= 3\chi \omega|E_\omega||E_{2\omega}| \sin(3\chi \omega t - \phi).
\end{split}
\end{equation}
Therefore, Eq.~\ref{eq:saddle_eqs} is satisfied for the times
\begin{equation}\label{eq:times}
t_n = \frac{\phi+n\pi}{3\chi \omega}, \quad n \text{ an integer.}
\end{equation}
We additionally require that the tunneling exponent is a minimum. Since the isotropic band separation increases with $|\mathbf{A}(t)|$, the simultaneous minima of $|\mathbf{A}(t)|$ and maxima of $|\mathbf{E}(t)|$ minimize $S_0(t)$. Thus, both equations in Eq.~\ref{eq:saddles} are satisfied when
\begin{equation}\label{eq:tunneling_times}
t_n^\text{min} = \frac{\phi+(2n+1)\pi}{3\chi \omega}, \quad n \text{ an integer.}
\end{equation}
The integer $n$ labels successive injection events with the same tunneling probability. Inserting these tunneling times into the trigonal-warping-corrected tunneling exponent in Eq.~\ref{eq:tunnel_exponent_warping}, we find
\begin{equation}
\begin{split}
    &\Delta_\tau(t_n^\text{min}) \simeq\\
    &=2 \sqrt{m^2+v_F^2\,|\tilde{A}|^2 + \frac{v_F^2\,a}{2} \tau \tilde{A}^3 \chi\cos \phi},
\end{split}
\end{equation}
with
\begin{equation}
    \tilde{A} = \frac{1}{\sqrt{2}\,{\omega}}\left( |E_\omega| - \frac{|E_{2\omega}|}{2}\right).
\end{equation}
Thus, we find that the trajectory that predominantly populates the valleys experiences different band separations for each valley, determined by the term $\tau\tilde{A}^3 \chi\cos\phi$, which is field-orientation dependent through $\phi$. 

Expanding to first order in the warping term correction and keeping the reduced electron-hole mass fixed at $\mu_0$, the difference in the tunneling exponent between the two valleys is
\begin{equation}\label{eq:diff_tunnel_exp}
\begin{split}
    &S_{K}(t_n^\text{min}) - S_{K'}(t_n^\text{min}) \propto \chi\tilde{A}^3\cos\phi.
\end{split}
\end{equation}
Since the tunneling rate $W_\tau(t_n)$ is exponentially suppressed by $S_\tau(t_n)$, this will produce a population imbalance between $K$ and $K'$ that is orientation dependent with $\phi$. In particular, if $S_K(t_n)>S_{K'}(t_n)$, then the valley populations at the end of the pulse $\rho_{K} < \rho_{K'}$, and viceversa.  Changing $\phi$ to $\phi+\pi$ exchanges the tunneling rates between the valleys and reverses the orientation-dependent contribution to the population imbalance.

The above expression is valid in the low-energy limit $a|\tilde{A}| \ll 1$ as long as the field has well-defined lobes that dominate the injection, i.e., that neither $|E_{\omega}| \to 0$ or $|E_{2\omega}|\to 0$. 

We note that the orientation-dependent warping term in the dispersion is still valid for graphene ($m=0$) which is in line with the observation and explanation in Mrudul et al.~\cite{mrudul2021controlling}. Similarly, for inversion-symmetric bilayer and bulk materials, where each pair of monolayers is equal but rotated by an angle $\pi$, the bicircular field has also been shown to produce orientation-dependent valley polarization. This is also in line with our result in the limit of weak interlayer coupling.

\subsection{General expression for two-color circular fields $\omega+N\omega$}
We can extend our derivation to a general combination of two circular frequencies $\omega$ and $N\omega$. When the two frequencies are counter-rotating, the fields contain $N+1$ lobes per cycle. When the two frequencies are co-rotating, then there are $N-1$ lobes per cycle.

\subsubsection{Counter-rotating two-color fields}
The general two-color counter-rotating field is
\begin{equation}\label{eq:efield_rotating_general}
    \mathbf{E}(t) = \text{Re}\left( |E_\omega|\,e^{- i \chi \omega t}\,\hat{\mathbf{e}}_+ + |E_{N\omega}|e^{i\phi}\,e^{-iN\chi\omega t}\,\hat{\mathbf{e}}_-\right).
\end{equation}
The times that minimize the isotropic tunneling exponent $S_0(t)$ in this case are
\begin{equation}
    t_{N,n}^\text{min} = \frac{\phi+(2n+1)\pi}{(N+1)\chi \omega}, \quad n \text{ an integer.}
\end{equation}
It follows that
\begin{equation}
\begin{split}
    &\Delta_\tau(t_{N,n}^\text{min}) \simeq\\
    &=2 \sqrt{m^2+v_F^2\,|\tilde{A}_N|^2 + \frac{v_F^2\,a}{2} \tau \tilde{A}_N^3 \chi\Gamma_{N}(n,\phi)},
\end{split}
\end{equation}
and
\begin{equation}\label{eq:general_tunnel_exp_counter}
    S_{K}(t_{N,n}^\text{min}) - S_{K'}(t_{N,n}^\text{min}) \propto \chi\tilde{A}_N^3\Gamma_N(n,\phi),
\end{equation}
where
\begin{equation}
    \tilde{A}_N = \frac{1}{\sqrt{2}\,{\omega}}\left( |E_\omega| - \frac{|E_{N\omega}|}{N}\right)
\end{equation}
and
\begin{equation}
    \Gamma_{N}(n,\phi) = -\sin \left[ \frac{3 (\phi+(2n+1)\pi)}{N+1} - \frac{3\pi}{2} \right].
\end{equation}
It is now easy to see the advantage of the bicircular field. Indeed, the warping contribution averaged over the dominant injection times from all lobes in one cycle is
\begin{equation}\label{eq:cycle-averaged-gamma}
    \sum_{n=0}^{N} \Gamma_{N}(n,\phi) = 0, \quad \text{for}\, N\neq 2.  
\end{equation}
For all other frequency combinations, the valley-dependent band separation contribution averages out over a cycle and thus there will be no field-orientation-dependence on the relative valley populations provided there are no envelope effects.

\subsubsection{Co-rotating two-color fields}
We define the co-rotating field as
\begin{equation}
    \mathbf{E} (t) = \text{Re}\left[ |E_\omega|e^{-i\chi\omega t}\,\hat{\mathbf{e}}_+  + |E_{N\omega}|e^{i(\phi+\pi)}e^{-iN\omega\chi t}\,\hat{\mathbf{e}}_+\right],
\end{equation}
where we have introduced a phase shift of \(\pi\) in the \(N\omega\)
field for later convenience. With this convention, the co-rotating \(\omega+4\omega\) field has the same \(\phi\)-dependent orientation factor as the counter-rotating \(\omega+2\omega\) field. We have that
\begin{equation}\label{eq:fields_squared_co}
    \begin{split}
        |\mathbf{E}(t)|^2 &= \frac{|E_\omega|^2+|E_{N\omega}|^2}{2} \\
        &- |E_\omega||E_{N\omega}|\cos((N-1)\chi\omega t - \phi),
    \end{split}
\end{equation}
\begin{equation}
    \begin{split}
        |\mathbf{A}(t)|^2 &= \frac{|E_\omega|^2+|E_{N\omega}|^2/N^2}{2\omega^2} \\
        &- \frac{|E_\omega||E_{N\omega}|}{N\omega^2}\cos((N-1)\chi\omega t - \phi).
    \end{split}
\end{equation}
In this case, we can find times that satisfy the extrema condition Eq.~\ref{eq:saddle_eqs} which are analogous to those in the counter-rotating case,
\begin{equation}
    t_{N,n} = \frac{\phi+(2n+1)\pi}{(N-1)\chi\omega}, \quad n \text{ an integer.}
\end{equation}
The main difference is that, unlike in the case of counter-rotating fields, here both $|\mathbf{E}(t)|$ and $|\mathbf{A}(t)|$ are maximum at the same times. Satisfying the minimum condition of Eq.~\ref{eq:saddles} additionally requires that
\begin{equation}
    \frac{3v_F^2 (|E_\omega|+|E_{N\omega}|)^2}{4N\omega^2m^2+2Nv_F^2 (|E_\omega|+|E_{N\omega}|/N)^2} < 1.
\end{equation}
Provided this condition is satisfied, $t_{N,n} = t_{N,n}^\text{min}$ and $S(t_{N,n}^\text{min})$ is a minimum. We then find similar expressions as for the counter-rotating case,
\begin{equation}\label{eq:zero-order-final-dispersion_general_co}
\begin{split}
    &\Delta_{\tau}(t_{N,n})
    =\\
    &=2
    \sqrt{m^2
    +
    v_F^2|\tilde{A}_{\text{co},N}|^2 +
    \frac{v_F^2a}{2}
    \tau
    \tilde{A}_{\text{co},N}^3
    \chi\Gamma_{\text{co},N}(n,\phi)}
\end{split}
\end{equation}
so that
\begin{equation}\label{eq:general_tunnel_exp_co}
    S_K(t_{N,n}) - S_{K'}(t_{N,n}) \propto \chi\tilde{A}_{\text{co},N}^3\Gamma_{\text{co},N}(n,\phi),
\end{equation}
where now
\begin{equation}
    \tilde{A}_{\text{co},N} = \frac{1}{\sqrt{2}\omega}\left( |E_\omega| + \frac{|E_{N\omega}|}{N}\right)
\end{equation}
and
\begin{equation}
    \Gamma_{\text{co},N}(n,\phi) = -\sin \left[ \frac{3 (\phi+(2n+1)\pi)}{N-1} - \frac{3\pi}{2} \right].
\end{equation}
In this case, we have that there are $N-1$ lobes per cycle, and
\begin{equation}
    \sum_{n=0}^{N-2} \Gamma_{\text{co},N}(n,\phi) = 0, \quad \text{for}\, N\neq 2,4.  
\end{equation}
For $N=2$, previous work has identified the existence of non-linear bulk photogalvanic currents~\cite{neufeld2021light}. These pulses display a highly circular Lissajous figure and thus we do not observe a switching of the valley polarization with the two-color phase $\phi$; the valley selectivity is dominated by the helicity. For $N=4$, recent work has also identified valley dichroism in graphene~\cite{PhysRevB.111.155421}. Indeed, for this configuration we obtain an equivalent expression as for the bicircular $\omega+2\omega$, albeit with a different effective vector potential $\tilde{A}_{\text{co},4}$.

\section{Numerical results}
In order to test our analytical expression, we run real-time simulations of the laser-crystal interaction using the reduced density matrix approach described in Molinero et al.~\cite{molinero2025semiconductor}. We use a full gapped graphene model, i.e., not limited to the low-energy approximation. We choose material parameters close to those of hexagonal boron nitride: $2m=5.9$~eV, $t=2.42$~eV, $a=1.44$~\AA. 

\begin{figure}
    \centering
        \includegraphics[width=\linewidth]{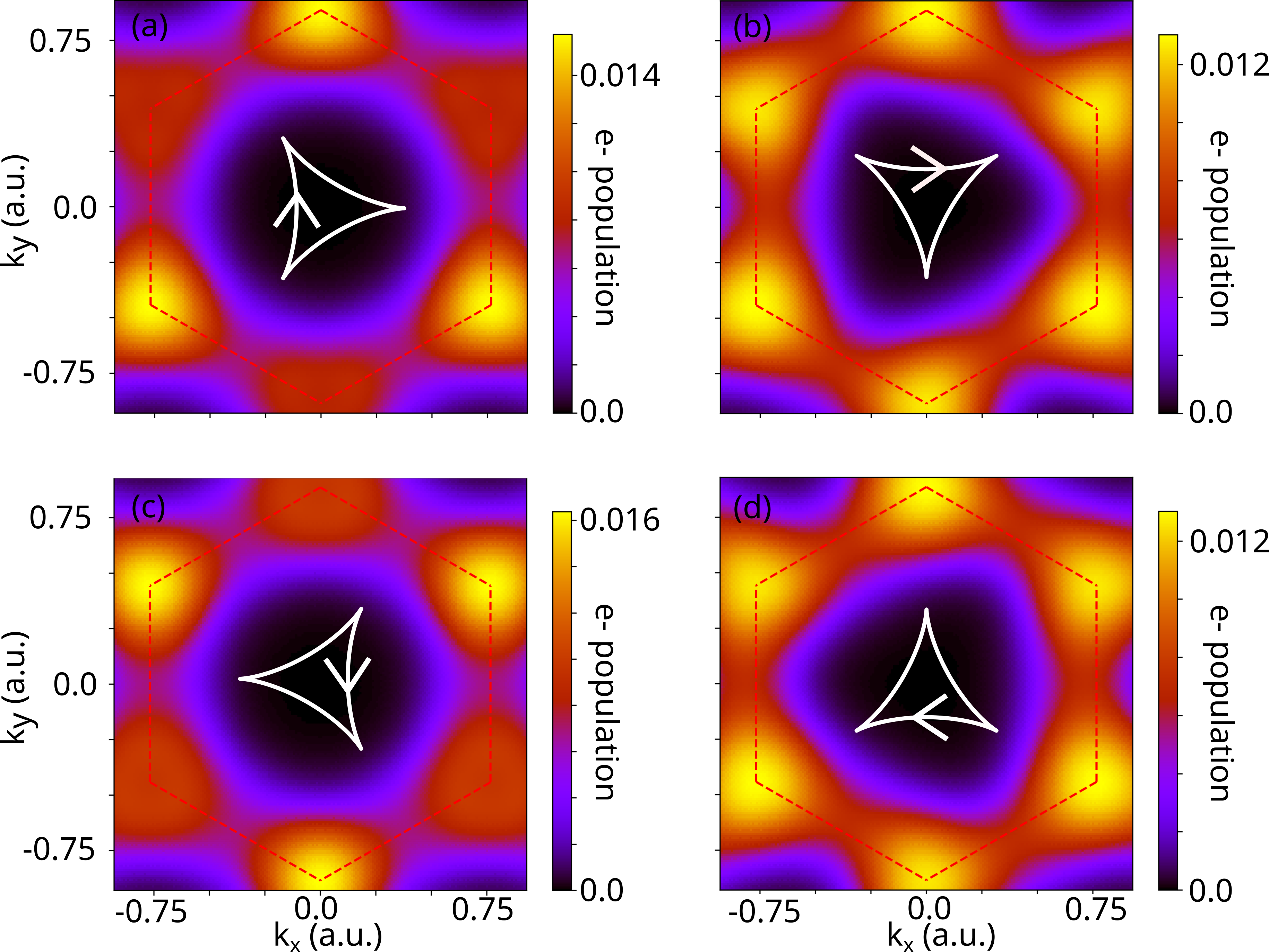}
    \caption{Valley populations in the conduction band after pulse interaction with a bicircular field defined by Eq.~\ref{eq:bicircular_field} with field strength ratio $r=0.5$ (see text). The first Brillouin zone is shown as the red hexagon, with the $K$ and $K'$ valleys at the vertices. The electric field is showed as the black trefoil in the center of the figures, with the arrow indicating the direction of rotation, which is always clockwise. Different values of the two-color phase $\phi$ are shown, controlling the orientation of the electric field relative to the lattice: (a) $\phi=0^\circ$, (b) $\phi=90^\circ$, (c) $\phi=180^\circ$, (d) $\phi=270^\circ$. }
    \label{fig:ratio_0_5}
\end{figure}

\begin{figure}
    \centering
        \includegraphics[width=\linewidth]{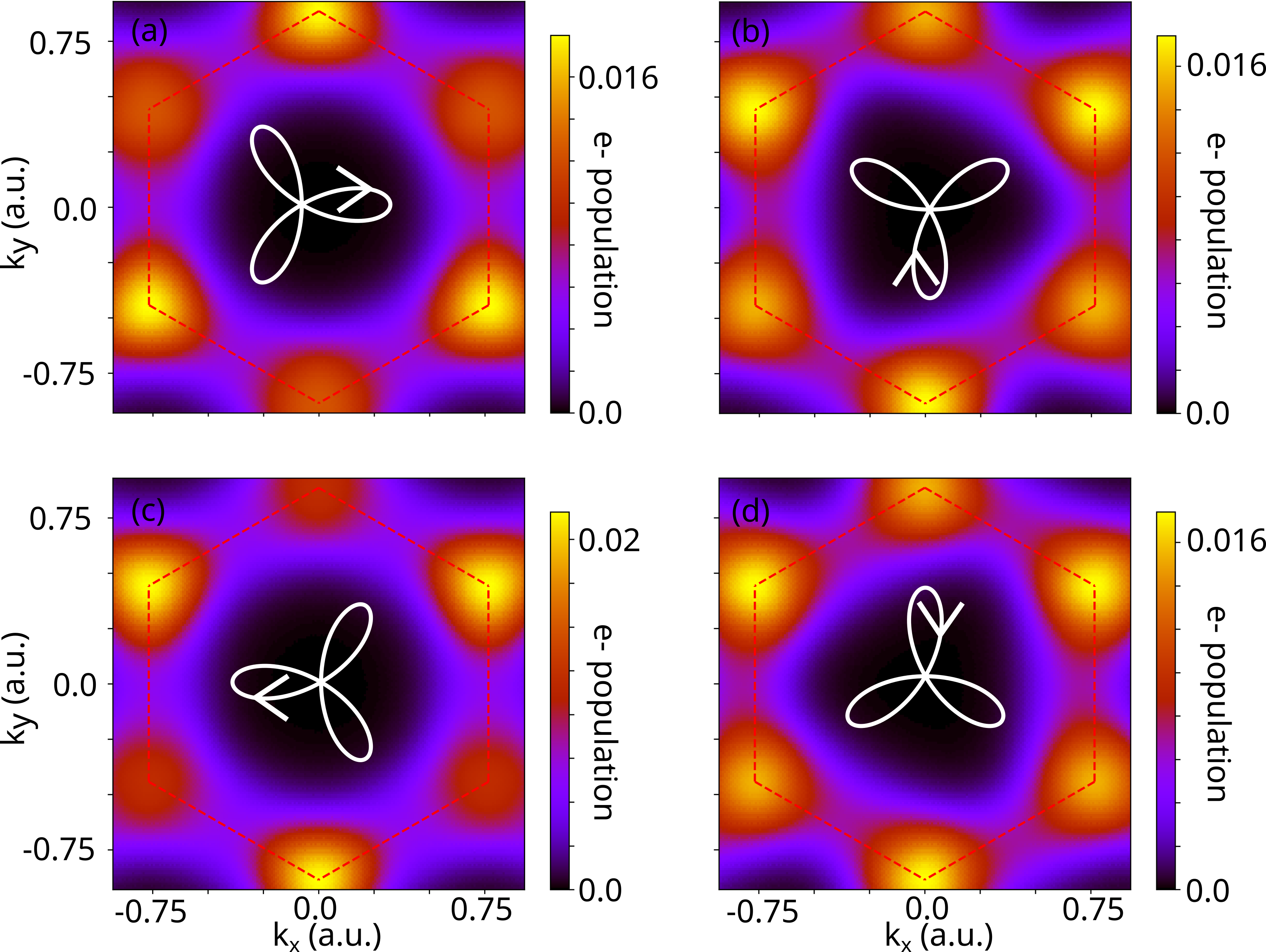}
    \caption{Same as Fig.~\ref{fig:ratio_0_5} for $r=1$.}
    \label{fig:ratio_1_0}
\end{figure}

\begin{figure}
    \centering
        \includegraphics[width=\linewidth]{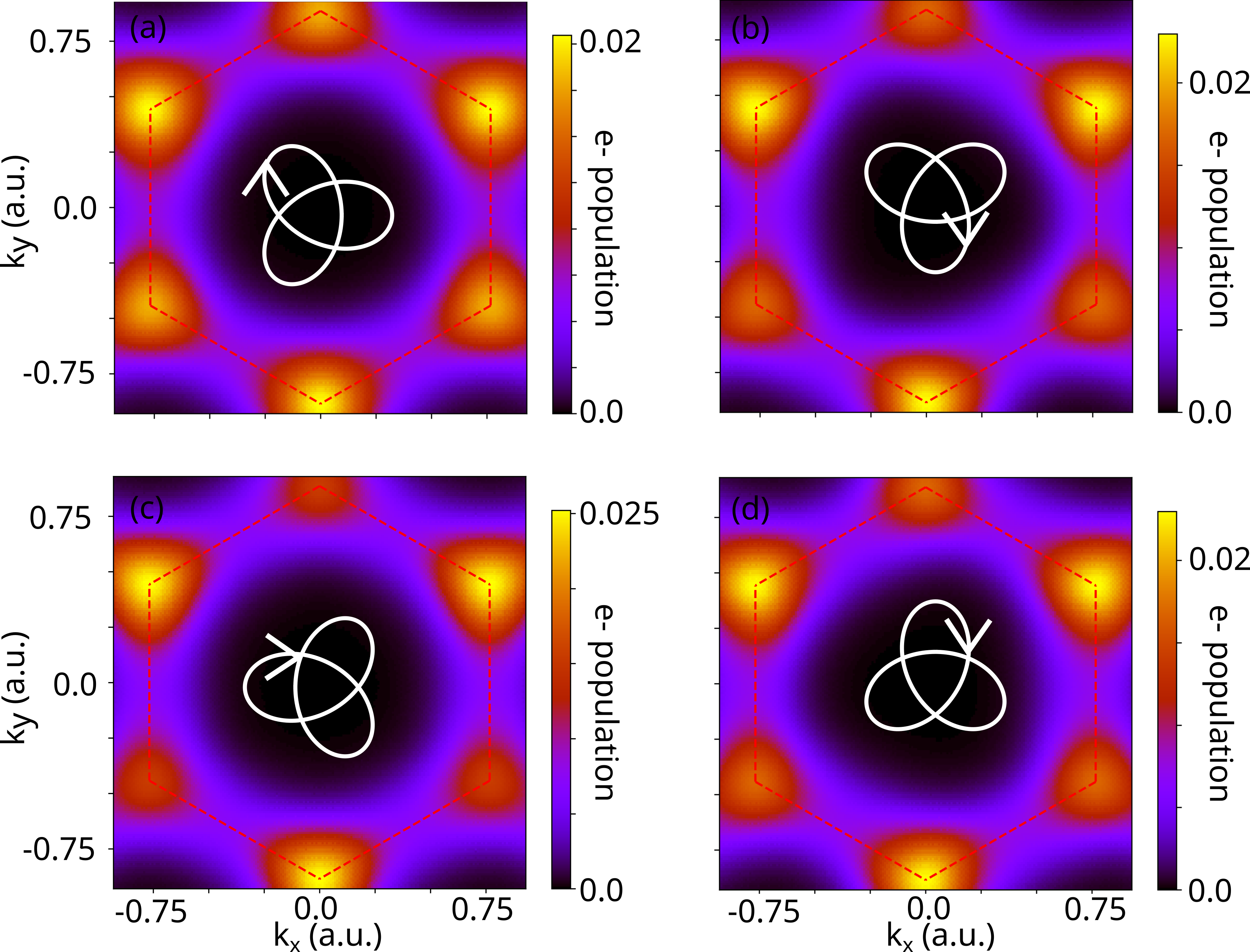}
    \caption{Same as Fig.~\ref{fig:ratio_0_5} for $r=2$.}
    \label{fig:ratio_2_0}
\end{figure}

We first consider a bicircular field defined by Eq.~\ref{eq:bicircular_field} with a driving frequency of $\omega=0.4$~eV and a flat-top envelope with 2 rising and falling cycles and 10 plateau cycles. The cycle-averaged intensity of the field is kept fixed at $I=|E_\omega|^2+|E_{2\omega}|^2 = 1$~TW/cm$^2$. This keeps the factor $a|\tilde{A}|\ll 1$ for all cases considered, so that the low-energy approximation is applicable. We also include a phenomenological dephasing $T_2=7$~fs, which corresponds to one cycle of the $\Omega=3\omega/2$ frequency. Different choices of dephasing time do not alter the main results. Figs.~\ref{fig:ratio_0_5}-~\ref{fig:ratio_2_0} show the electron population in the conduction band after the interaction with the pulse for different values of the ratio $r=\frac{|E_{2\omega|}}{|E_\omega|}$ and the two-color phase delay $\phi$, for a fixed helicity $\chi=+1$. 

When $r<2$ we have that $\tilde{A} > 0$ and Eqs.~\ref{eq:diff_tunnel_exp} predicts that $\rho_K < \rho_{K'}$ for $\phi=0$ and$\rho_k > \rho_{K'}$ for $\phi=\pi$. This is confirmed by the numerical calculations in panels a,c of Figs.~\ref{fig:ratio_0_5},\ref{fig:ratio_1_0}. 

When $\phi=\pi/2,3\pi/2$, the trigonal warping term vanishes and the valley polarization is dominated by the helicity of the fields. For a ratio of $r=0.5$, the fundamental field is stronger than the second harmonic, and there is a slight valley polarization of $K'$ (Fig.~\ref{fig:ratio_0_5}b,d), corresponding to that favored by the fundamental field's helicity. In contrast, when $r=1.0$, the favored valley changes to $K$ (Fig.~\ref{fig:ratio_1_0}b,d), which is coupled to the helicity of the $2\omega$ field.

When $r=2$, then $\tilde{A}=0$ and the orientation-dependence $\phi$ in Eq.~\ref{eq:diff_tunnel_exp} vanishes. The valley polarization reduces to only the helicity-dependent contribution, and is therefore the same for all orientations. This is also confirmed by the numerical simulations in Fig.~\ref{fig:ratio_2_0}a-d, where the same valley $K$ is populated irrespective of $\phi$. 

For $r>2$, then $\tilde{A}<0$ and the valley asymmetry should switch relative to the case for $r<2$ according to Eq.~\ref{eq:diff_tunnel_exp}. We do not observe this switch in the numerical simulations, which display the same behavior as in Fig.~\ref{fig:ratio_2_0} and thus are not shown. The absence of the switch is most likely because for $r>2$ the pulse has very broad lobes, almost resembling a circular pulse, and thus the single tunneling time approximation fails.


\begin{figure}
    \centering
        \includegraphics[width=\linewidth]{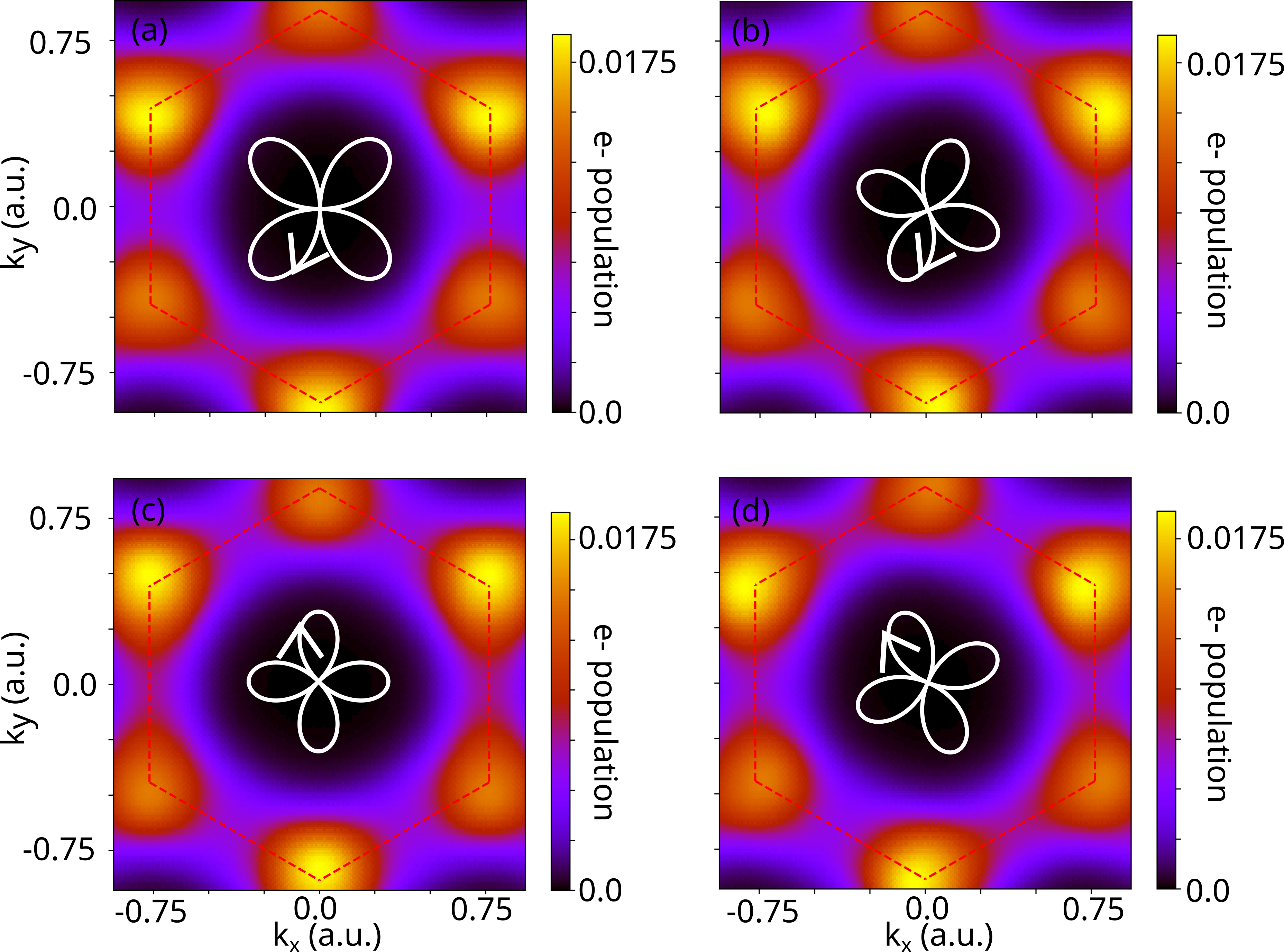}
    \caption{Valley populations in the conduction band after pulse interaction with a counter-rotating $\omega+3\omega$ field with field strength ratio $r=1.0$. The helicity of the field is always clockwise. Different values of the two-color phase $\phi$ are shown, controlling the orientation of the electric field relative to the lattice: (a) $\phi=0^\circ$, (b) $\phi=90^\circ$, (c) $\phi=180^\circ$, (d) $\phi=270^\circ$. }
    \label{fig:ratio_1_0_3w}
\end{figure}

In Fig.~\ref{fig:ratio_1_0_3w} we show results for a pulse $\omega+3\omega$ with ratio $r=\frac{|E_{3\omega}|}{|E_\omega|}=1$. The Lissajous figure of the field is made of four lobes, as we show in the insets. In this case, the orientation-dependent contribution to the valley bandgap vanishes per cycle (see Eq.~\ref{eq:cycle-averaged-gamma}) and thus the only valley-asymmetric contribution is the helicity-dependent contribution, again favoring $K$ for all $\phi$.

\begin{figure}
    \centering
        \includegraphics[width=\linewidth]{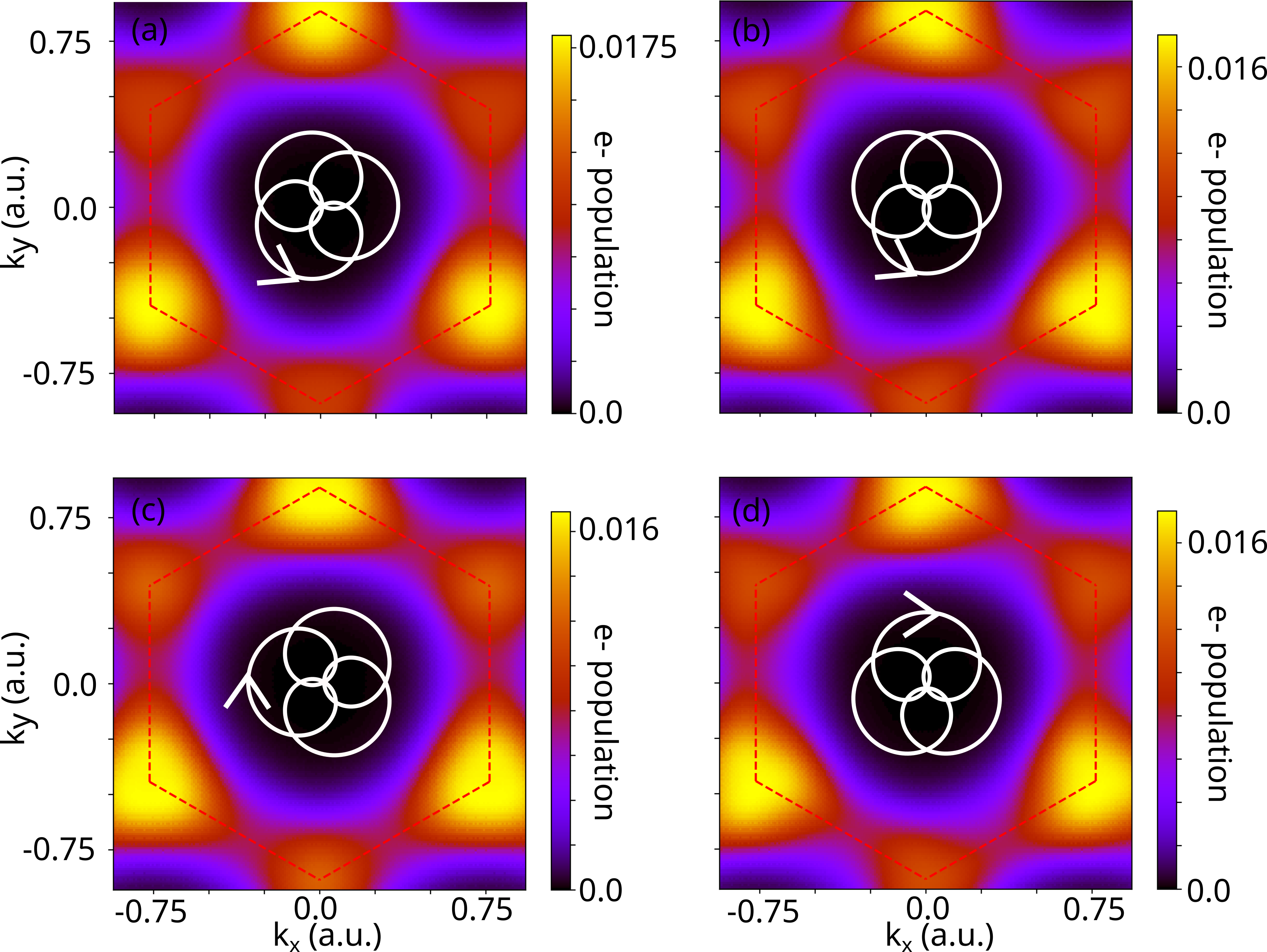}
    \caption{Valley populations in the conduction band after pulse interaction with a co-rotating $\omega+4\omega$ field with field strength ratio $r=1.0$. The helicity of the field is always counter-clockwise. Different values of the two-color phase $\phi$ are shown, controlling the orientation of the electric field relative to the lattice: (a) $\phi=0^\circ$, (b) $\phi=90^\circ$, (c) $\phi=180^\circ$, (d) $\phi=270^\circ$. }
    \label{fig:ratio_1_0_co}
\end{figure}

\begin{figure}
    \centering
        \includegraphics[width=\linewidth]{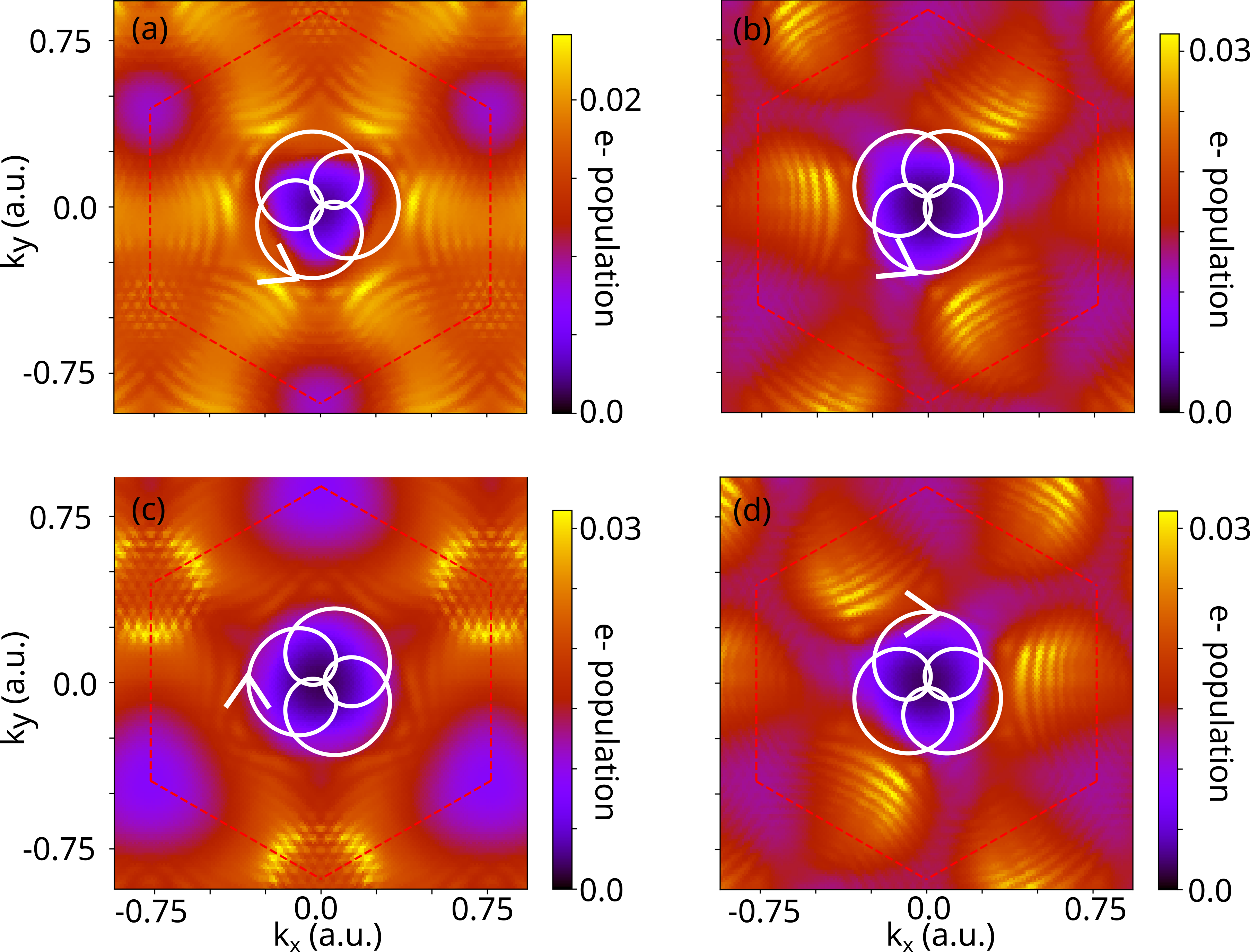}
    \caption{Same as Fig.~\ref{fig:ratio_1_0_co} but for an intensity four times larger, $I=4$~TW/cm$^2$, now showing valley switching with the field orientation $\phi$.}
    \label{fig:ratio_1_0_co_highint}
\end{figure}

We last calculate the electron populations for co-rotating $\omega+4\omega$ field in Fig.~\ref{fig:ratio_1_0_co} and Fig.~\ref{fig:ratio_1_0_co_highint}. The Lissajous figure resembles the trefoil pattern of the counter-rotating $\omega+2\omega$ field, albeit with wider lobes for the ratio $r=1$ considered. Using the same intensity as before, $I=|E_\omega|^2+|E_{4\omega}|^2 = 1$TW/cm$^2$, the valley polarization does not switch as a function of $\phi$. The reason is that, for the co-rotating case, both fields share the same helicity and thus reinforce the helicity-dependent valley physics. The field-orientation term is in this case not strong enough to counter the helicity valley selection. However, as we increase the intensity to $I=|E_\omega|^2+|E_{4\omega}|^2 = 4$TW/cm$^2$ in Fig.~\ref{fig:ratio_1_0_co_highint}, and thus increase the weight of the field-orientation term, we do observe the valley switching as a function of $\phi$, as Eq.~\ref{eq:general_tunnel_exp_co} predicts, as in the case of the bicircular field. However, at $4$~TW/cm$^2$, the low-energy cubic expansion is no longer controlled since $a|\tilde{A}_\text{co}| \simeq 1.24$, so this comparison should be regarded as qualitative.

\section{Conclusions\protect\label{sec:conclusions}}
We have derived an analytical expression for the effective bandgap of hexagonal systems during the tunneling event under strong two-color circular fields. The strong-field valley-resolved gap, whose leading valley-dependent contribution arises from the trigonal warping term, leads to field-orientation-dependence of the valley polarization. Our derivation also demonstrates that this dependence is only relevant for trigonally-symmetric fields: either counter-rotating $\omega+2\omega$ or co-rotating $\omega+4\omega$. 

We have successfully tested the predictions of our analytical expression with numerical calculations. In addition, we have derived the helicity-dependent optical selection rules by using a rotating-frame transformation. Our derivation offers a complementary approach to the observed valley polarization in trefoil-symmetric fields that is based on the effective tunneling bandgap experienced by the trajectory that leads to $K$ and $K'$ populations. This derivation is applicable to inversion symmetric bilayers and bulk systems in the 2H phase, provided the interlayer coupling is weak.

\section*{Acknowledgments}

This work was supported through the Spanish Ministry of Science, Innovation and Universities through grant references CNS2025-166331, PID2023-146676NA-I00, CNS2024-154463 and RYC2022-035373-I, funded by MICIU/AEI/10.13039/501100011033, “ERDF A way of making Europe” and “ESF+”, and the Atracci\'on de Talento Comunidad de Madrid Fellowship 2022-T1/IND24102.

\section*{Appendix}

\subsection{Low-energy approximation of gapped graphene}\label{sec:low-energy}
We consider a gapped graphene system with sublattices A and B, with an on-site energy $m$ on A sites and $-m$ on B sites. We include only real first-neighbour hoppings between opposite sublattices and no second-neighbour hopping term. The Hamiltonian can be written as~\cite{castro2009electronic}
\begin{equation}
H_0(\mathbf k)
=
\begin{pmatrix}
m & h(\mathbf{k}) \\
h^*(\mathbf k) & -m
\end{pmatrix},
\end{equation}
with
\begin{equation}
    h(\mathbf{k}) = -t f(\mathbf{k}), \quad
    f(\mathbf k)
=
\sum_{j=1}^{3}
e^{i\mathbf k\cdot\boldsymbol\delta_j},
\end{equation}
and $\delta_j$ the first-neighbour vectors, which we choose as
\begin{equation}
\begin{split}
&\boldsymbol\delta_1=a\left(\frac{1}{2},\frac{\sqrt{3}}{2}\right),\\
&\boldsymbol\delta_2=a\left(\frac{1}{2},-\frac{\sqrt{3}}{2}\right),\\
&\boldsymbol\delta_3=a (-1,0).
\end{split}
\end{equation}
Above, $a$ is the nearest-neighbour distance, such that the lattice constant $a_\text{latt} = \sqrt{3}\,a$. The Hamiltonian can be written in terms of Pauli matrices,
\begin{equation}
H_0(\mathbf k)
=
d_x(\mathbf k)\sigma_x
+
d_y(\mathbf k)\sigma_y
+
m\sigma_z,
\end{equation}
where
\begin{equation}
d_x(\mathbf k)=-t\,\mathrm{Re}\,f(\mathbf k),
\qquad
d_y(\mathbf k)=t\,\mathrm{Im}\,f(\mathbf k).
\end{equation}

The band separation is thus
\begin{equation}
\Delta(\mathbf{k}) = 2\sqrt{m^2 + 
|h(\mathbf{k})|^2}
.
\end{equation}

The two inequivalent valleys are located at
\begin{equation}
    \mathbf{K}_\tau = \left( \frac{2\pi}{3a}, \tau\frac{2\pi}{3\sqrt{3}a}\right)
\end{equation}
where $\tau=\pm 1$ specifies the valley index (+1 for $\mathbf{K}$ and -1 for $\mathbf{K}'$). At the two valleys, one has \(h(\mathbf{K}_\tau)=0\), and therefore the field-free direct band gap at both valleys is
\begin{equation}
\Delta (\mathbf{K}_\tau) = 2|m|.
\end{equation}

Since we are interested in valley physics, we adopt the common low-energy approximation by expanding $h(\mathbf{k})$ around a valley, \(\mathbf k=\mathbf K_\tau+\mathbf q\)~\cite{castro2009electronic,xiao2007valley}. To first order in the Taylor expansion,
\begin{equation}
    h_\tau^{(1)}(\mathbf{q})=
v_Fe^{i\varphi}
\left(q_x+i\tau q_y\right) + O(|\mathbf{q}|^2),
\end{equation}
where $v_F = \frac{3ta}{2}$. The phase $\varphi=-\pi/6$ depends on our lattice convention and will not be relevant, but we keep it for completeness. In polar coordinates, 
\begin{equation}
    q_{x}=|\mathbf{q}|\cos\theta_q,
    \qquad
    q_{y}=|\mathbf{q}|\sin\theta_q,
\end{equation}
where
\begin{equation}
\begin{split}
    &|\mathbf{q}|^2 = |q_x|^2 + |q_y|^2,\\
    &\theta_q = \arg \left\{ q_x + i q_y \right\}.
\end{split}
\end{equation}
we have
\begin{equation}
    h_\tau^{(1)}(\mathbf{q}) = v_F e^{i\varphi} |\mathbf{q}| e^{i \tau \theta_q}.
\end{equation}
To first order, the low-energy Hamiltonian is isotropic and its band separation is
\begin{equation}\label{eq:sq_energy_isotropic}
\begin{split}
    \Delta_0(|\mathbf{q}|) &= 2 \sqrt{m^2 + |h_\tau^{(1) }(\mathbf{q})|^2} \\
    &=2\sqrt{m^2 + v_F^2\,|\mathbf{q}|^2)}.
\end{split}
\end{equation}
The reduced electron-hole mass $\mu_0$ of the isotropic first-order Hamiltonian can be obtained by identifying the parabolic band energy
\begin{equation}
    \Delta_0(|\mathbf{q}|) = 2m+\frac{|\mathbf{q}|^2}{2\mu_0}
\end{equation}
with Eq.~\ref{eq:sq_energy_isotropic}, and expanding for $v_F |\mathbf{q}| \ll m$,
\begin{equation}
    \Delta_0(|\mathbf{q}|) = 2m \sqrt{1+\frac{v_F^2\,|\mathbf{q}|^2}{m^2}} \approx 2m + \frac{v_F^2\,|\mathbf{q}|^2}{m},
\end{equation}
so that
\begin{equation}\label{eq:eh_mass}
    \mu_0 = \frac{m}{2v_F^2}.
\end{equation}

The first order is isotropic and thus not enough to distinguish the valleys. We therefore include the next term in the expansion of $h(\mathbf{k})$,
\begin{equation}
\begin{split}
h_\tau^{(2)} (\mathbf{q}) =
-\frac{i v_F a}{4}e^{i\varphi}|\mathbf{q}|^2 e^{-2i\tau\theta_q} + O(|\mathbf{q}|^3).
\end{split}
\end{equation}
The band separation of the second-order low-energy Hamiltonian is then
\begin{equation}\label{eq:static_energy_polar}
\begin{split}
&\Delta_{\tau}(\mathbf{q}) = 2\sqrt{m^2+|h_\tau^{(1)}(\mathbf{q}) + h_\tau^{(2)}(\mathbf{q})|^2} \\
    &= 2\sqrt{m^2+v_F^2\,|\mathbf{q}|^2 - \frac{v_F^2 a}{2}\tau |\mathbf{q}|^3\sin 3\theta_q}.
\end{split}
\end{equation}
The third term is the non-isotropic trigonal warping term, which now contains the valley index $\tau$.

In the presence of a time-dependent field described by a vector potential $\mathbf{A}(t) = \{A_x(t),A_y(t)\}$, we replace
\begin{equation}
    q_x \to q_x(t) = q_x+A_x(t), \quad q_y\to q_y(t) = q_y+A_y(t),
\end{equation}
so that the band separation in Eq.~\ref{eq:static_energy_polar} becomes
\begin{equation}\label{eq:second_order_dispersion_lab_frame}
\begin{split}
    &\Delta_\tau(\mathbf{q},t) \\
    &=2\sqrt{ m^2+v_F^2 |\mathbf{q}(t)|^2 - \frac{v_F^2 a}{2}\tau |\mathbf{q}(t)|^3\sin3\theta_{q(t)}}.
\end{split}
\end{equation}

Finally, since we are interested in the trajectory that ends at the valley center $\mathbf{q}=0$, then $\mathbf{q}(t)=\mathbf{A}(t)$ and
\begin{equation}
\begin{split}
    &\Delta_\tau(t)\\
    &=2 \sqrt{m^2+v_F^2\,|\mathbf{A}(t)|^2 - \frac{v_F^2\,a}{2} \tau |\mathbf{A}(t)|^3 \sin 3\theta_{A(t)}}.
\end{split}
\end{equation}

\subsection{Helicity-dependent optical valley selection rules from a rotating frame perspective}~\label{sec:helicity}
We consider weak-field, one-photon transitions at the valley centers and thus only retain the linear, isotropic expansion of the Hamiltonian $H_\tau(\mathbf{q};t)$. At $\mathbf{q}=0$, this Hamiltonian is
\begin{equation}
H_\tau(t)=
\begin{pmatrix}
m & h^{(1)}_\tau(t)\\
h_\tau^{(1)*}(t)&-m
\end{pmatrix},
\end{equation}
with
\begin{equation}
h_\tau^{(1)}(t)=v_Fe^{i\varphi}[A_x(t)+i\tau A_y(t)].    
\end{equation}
Following the idea of previous works~\cite{reich2016rotating, reich2016illuminating, pisanty2017strong}, we introduce a valley-dependent rotation transformation of the sublattice pseudospin $\tau\sigma_z/2$,
\begin{equation}\label{eq:unitary_matrix}
\begin{split}
&|\psi^R(t)\rangle
=
U_\tau(t)|\psi(t)\rangle,\\
&U_\tau(t)
=
e^{i\frac{\alpha \tau}{2}t\,\sigma_z},
\end{split}
\end{equation}
where $\alpha$ is the rotating frequency. The Hamiltonian then transforms as
\begin{equation}
\begin{split}
H_\tau^R(t)
&=
U_\tau(t)H_\tau(t)U_\tau^\dagger(t)
+
i\dot U_\tau(t)U_\tau^\dagger(t)\\
&=H_{\tau,0}^{R} + V_\tau^R(t),
\end{split}
\end{equation}
where
\begin{equation}
    \begin{split}
        &H_{\tau,0}^R = \left(m-\frac{\alpha\tau}{2}\right) \sigma_z, \\
        &V_{\tau}^R(t) = \begin{pmatrix}
    0 &  h_\tau^{(1)R}(t) \\
    h_\tau^{*(1)R}(t) & 0,
\end{pmatrix}
    \end{split}
\end{equation}
and
\begin{equation}
    h_\tau^{(1)R}(t) = e^{i\alpha\tau t} h_\tau^{(1)}(t).
\end{equation}
The first-order probability amplitude from an initially occupied valence state to a conduction state is
\begin{equation}
    \mathcal{A}_{c\gets v}^{(1)} (t) = -i \int_{-\infty}^t dt' e^{i (2m-\alpha\tau) t'} h_\tau^{(1) R}(t').
\end{equation}
From Eq.~\ref{eq:bicircular_vpot},
\begin{equation}
\begin{split}
    &A_x(t) + i\tau A_y(t) \\
    &= -\frac{i\chi \tau}{\sqrt{2}\omega} \left[ |E_\omega| e^{i\chi\tau\omega t} + \frac{|E_{2\omega}|}{2}e^{i\tau \phi}e^{-2i\chi\tau\omega t}\right],
\end{split}
\end{equation}
and therefore
\begin{equation}
\begin{split}
    h_\tau^{(1)R}(t) &= -\frac{i\chi\tau v_F e^{i\varphi}}{\sqrt{2}\omega}\Bigg[ |E_\omega| e^{i\tau (\chi\omega+\alpha) t} \\
    &+ \frac{|E_{2\omega}|}{2}e^{i\tau \phi}e^{-i\tau(2\chi\omega - \alpha) t}\Bigg].
\end{split}
\end{equation}
As in \cite{pisanty2017strong}, we now choose
\begin{equation}
\alpha\equiv\frac{\chi\omega}{2},
\end{equation}
which shifts both colors to the same rotating-frame frequency magnitude
\begin{equation}
\Omega=\frac{3\omega}{2},
\end{equation}
such that
\begin{equation}
    h_\tau^{(1)R}(t) = -\frac{i\chi\tau v_F e^{i\varphi}}{\sqrt{2}\omega}\left[ |E_\omega| e^{i\chi\tau\Omega t} + \frac{|E_{2\omega}|}{2}e^{i\tau \phi}e^{-i\chi\tau\Omega t}\right].
\end{equation}
The first order probability amplitude is then
\begin{equation}
\begin{split}
    \mathcal{A}_{c\gets v}^{(1)} (t) &= -\frac{\chi\tau v_F e^{i\varphi}}{\sqrt{2}\omega} \Bigg[ |E_\omega| \int_{-\infty}^t dt' e^{i (2m+\chi\tau\omega) t'} \\
    &+ \frac{|E_{2\omega}|}{2} e^{i\tau\phi}  \int_{-\infty}^t dt' e^{i (2m-2\chi\tau\omega) t'} \Bigg].
\end{split}
\end{equation}
Taking the monochromatic limit $t\to \infty$, we find that
\begin{equation}\label{eq:one_photon_valley_rotating}
\begin{split}
    \mathcal{A}_{c\gets v}^{(1)} (t) &= -\frac{2\pi\chi\tau v_F e^{i\varphi}}{\sqrt{2}\omega} \Bigg[ |E_\omega| \delta (2m+\chi\tau\omega) \\
    &+ \frac{|E_{2\omega}|}{2} e^{i\tau\phi}  \delta (2m-2\chi\tau\omega) \Bigg].
\end{split}
\end{equation}
When the fundamental component is gap-resonant, $\omega=2m$, the valley for which one-photon excitation is allowed is that for which $\chi\tau=-1$ (first term in Eq.~\ref{eq:one_photon_valley_rotating}). When the second harmonic is gap-resonant, $2\omega=2m$, then the excited valley is the opposite, $\chi\tau=+1$ (second term in Eq.~\ref{eq:one_photon_valley_rotating}). These are the one-photon valley selection rules~\cite{xiao2007valley}, expressed in a pseudospin rotating frame where both driving components of the bicircular field have the same frequency magnitude.

\subsection{First order trigonal warping correction to the tunneling rate}\label{sec:warping_correction}
In the low-energy approximation, the trigonal warping term is the second order correction to the Hamiltonian, with the isotropic contribution being the first order correction (see Section II). For the isotropic tunneling exponent $S_0(t)$, the dominant injection time $t_n$ satisfies
\begin{equation}
    \partial_t S_0(t)\Big|_{t=t_n} = 0.
\end{equation}
Including the second-order warping term contribution changes both the exponent and the optimal injection times,
\begin{equation}
    S(t) = S_0(t) + \delta S(t), \quad t_n \to t_n + \delta t.
\end{equation}
We now expand around that time correction up to second order,
\begin{equation}
\begin{split}
    S(t_n+\delta t) &= S_0(t_n) + \delta S(t_n) + \frac{\partial{S_0(t)}}{\partial t}\Big|_{t=t_n} \delta t \\
    &+ \frac{\partial({\delta S(t)})}{\partial t}\Big|_{t=t_n} \delta t + \frac{1}{2}\frac{\partial^2{S_0(t)}}{\partial t^2}\Big|_{t=t_n} \delta t^2 \\
    &+ O(\delta^3).
\end{split}
\end{equation}
The last two terms are quadratic on the perturbation $\delta$. The third term is zero at the saddle, $\frac{d S_0(t)}{dt}|_{t=t_n} = 0$. Therefore, inserting the saddle times found with the isotropic Hamiltonian into the trigonal-warping-corrected tunneling exponent is correct up to first order in the trigonal warping perturbation.

\bibliographystyle{apsrev4-1}
\bibliography{biblio}

@article{vitale2018,
author = {Vitale, Steven A. and Nezich, Daniel and Varghese, Joseph O. and Kim, Philip and Gedik, Nuh and Jarillo-Herrero, Pablo and Xiao, Di and Rothschild, Mordechai},
title = {Valleytronics: Opportunities, Challenges, and Paths Forward},
journal = {Small},
volume = {14},
number = {38},
pages = {1801483},
doi = {https://doi.org/10.1002/smll.201801483},
url = {https://onlinelibrary.wiley.com/doi/abs/10.1002/smll.201801483},
year = {2018}
}

@article{schaibley2016valleytronics,
  title={Valleytronics in 2D materials},
  author={Schaibley, John R and Yu, Hongyi and Clark, Genevieve and Rivera, Pasqual and Ross, Jason S and Seyler, Kyle L and Yao, Wang and Xu, Xiaodong},
  journal={Nature Reviews Materials},
  volume={1},
  number={11},
  pages={16055},
  year={2016},
  publisher={Nature Publishing Group}
}

@article{mak2012control,
  title={Control of valley polarization in monolayer MoS2 by optical helicity},
  author={Mak, Kin Fai and He, Keliang and Shan, Jie and Heinz, Tony F},
  journal={Nature Nanotechnology},
  volume={7},
  number={8},
  pages={494--498},
  year={2012},
  publisher={Nature Publishing Group UK London}
}

@article{xiao2007valley,
  title={Valley-Contrasting Physics in Graphene: Magnetic Moment and Topological Transport},
  author={Xiao, Di and Yao, Wang and Niu, Qian},
  journal={Physical Review Letters},
  volume={99},
  number={23},
  pages={236809},
  year={2007},
  publisher={APS}
}

@article{yao2008valley,
  title={Valley-dependent optoelectronics from inversion symmetry breaking},
  author={Yao, Wang and Xiao, Di and Niu, Qian},
  journal={Physical Review B—Condensed Matter and Materials Physics},
  volume={77},
  number={23},
  pages={235406},
  year={2008},
  publisher={APS}
}

@article{xiao2010berry,
  title={Berry phase effects on electronic properties},
  author={Xiao, Di and Chang, Ming-Che and Niu, Qian},
  journal={Reviews of Modern Physics},
  volume={82},
  number={3},
  pages={1959--2007},
  year={2010},
  publisher={APS}
}

@article{hao2016direct,
  title={Direct measurement of exciton valley coherence in monolayer WSe2},
  author={Hao, Kai and Moody, Galan and Wu, Fengcheng and Dass, Chandriker Kavir and Xu, Lixiang and Chen, Chang-Hsiao and Sun, Liuyang and Li, Ming-Yang and Li, Lain-Jong and MacDonald, Allan H and others},
  journal={Nature Physics},
  volume={12},
  number={7},
  pages={677--682},
  year={2016},
  publisher={Nature Publishing Group UK London}
}

@Article{Gucci2026,
author={Gucci, Francesco
and Molinero, Eduardo B.
and Russo, Mattia
and San-Jose, Pablo
and Camargo, Franco V. A.
and Maiuri, Margherita
and Ivanov, Misha
and Jim{\'e}nez-Gal{\'a}n, {\'A}lvaro
and Silva, Rui E. F.
and Dal Conte, Stefano
and Cerullo, Giulio},
title={Encoding and manipulating ultrafast coherent valleytronic information with lightwaves},
journal={Nature Photonics},
year={2026},
month={Mar},
day={01},
volume={20},
number={3},
pages={266-272},
issn={1749-4893},
doi={10.1038/s41566-025-01823-w},
url={https://doi.org/10.1038/s41566-025-01823-w}
}

@article{ye2017optical,
  title={Optical manipulation of valley pseudospin},
  author={Ye, Ziliang and Sun, Dezheng and Heinz, Tony F},
  journal={Nature Physics},
  volume={13},
  number={1},
  pages={26--29},
  year={2017},
  publisher={Nature Publishing Group UK London}
}

@article{jimenez2021sub,
  title={Sub-cycle valleytronics: control of valley polarization using few-cycle linearly polarized pulses},
  author={Jim{\'e}nez-Gal{\'a}n, {\'A}lvaro and Silva, Rui EF and Smirnova, Olga and Ivanov, Misha},
  journal={Optica},
  volume={8},
  number={3},
  pages={277--280},
  year={2021},
  publisher={Optical Society of America}
}

@article{jimenez2020lightwave,
  title={Lightwave control of topological properties in 2D materials for sub-cycle and non-resonant valley manipulation},
  author={Silva, REF and Smirnova, O and Ivanov, M},
  journal={Nature Photonics},
  volume={14},
  number={12},
  pages={728--732},
  year={2020},
  publisher={Nature Publishing Group UK London}
}

@article{mrudul2021light,
  title={Light-induced valleytronics in pristine graphene},
  author={Mrudul, MS and Jim{\'e}nez-Gal{\'a}n, {\'A}lvaro and Ivanov, Misha and Dixit, Gopal},
  journal={Optica},
  volume={8},
  number={3},
  pages={422--427},
  year={2021},
  publisher={Optical Society of America}
}

@article{mitra2024light,
  title={Light-wave-controlled Haldane model in monolayer hexagonal boron nitride},
  author={Mitra, Sambit and Jim{\'e}nez-Gal{\'a}n, {\'A}lvaro and Aulich, Mario and Neuhaus, Marcel and Silva, Rui EF and Pervak, Volodymyr and Kling, Matthias F and Biswas, Shubhadeep},
  journal={Nature},
  volume={628},
  number={8009},
  pages={752--757},
  year={2024},
  publisher={Nature Publishing Group UK London}
}

@article{tyulnev2024valleytronics,
  title={Valleytronics in bulk MoS2 with a topologic optical field},
  author={Tyulnev, Igor and Jim{\'e}nez-Gal{\'a}n, {\'A}lvaro and Poborska, Julita and Vamos, Lenard and Russell, Philip St J and Tani, Francesco and Smirnova, Olga and Ivanov, Misha and Silva, Rui EF and Biegert, Jens},
  journal={Nature},
  volume={628},
  number={8009},
  pages={746--751},
  year={2024},
  publisher={Nature Publishing Group UK London}
}

@article{mrudul2021controlling,
  title={Controlling valley-polarisation in graphene via tailored light pulses},
  author={Mrudul, MS and Dixit, Gopal},
  journal={Journal of Physics B: Atomic, Molecular and Optical Physics},
  volume={54},
  number={22},
  pages={224001},
  year={2021},
  publisher={IOP Publishing}
}

@article{oka2009photovoltaic,
  title={Photovoltaic Hall effect in graphene},
  author={Oka, Takashi and Aoki, Hideo},
  journal={Physical Review B—Condensed Matter and Materials Physics},
  volume={79},
  number={8},
  pages={081406},
  year={2009},
  publisher={APS}
}

@article{mciver2020light,
  title={Light-induced anomalous Hall effect in graphene},
  author={McIver, JW and Schulte, B and Stein, F-U and Matsuyama, T and Jotzu, G and Meier, G and Cavalleri, A},
  journal={Nature Physics},
  volume={16},
  number={1},
  pages={38--41},
  year={2020},
  publisher={Nature Pub. Group}
}

@incollection{molinero2024manipulation,
  title={Manipulation of Quantum Properties in 2D Materials with Strong Tailored Fields},
  author={Molinero, Eduardo B and Silva, Rui EF and Jim{\'e}nez-Gal{\'a}n, {\'A}lvaro},
  booktitle={High-Order Harmonic Generation in Solids},
  pages={17--49},
  year={2024},
  publisher={World Scientific}
}

@article{pisanty2017strong,
  title={Strong-field approximation in a rotating frame: High-order harmonic emission from p states in bicircular fields},
  author={Pisanty, Emilio and Jim{\'e}nez-Gal{\'a}n, {\'A}lvaro},
  journal={Physical Review A},
  volume={96},
  number={6},
  pages={063401},
  year={2017},
  publisher={APS}
}

@article{castro2009electronic,
  title = {The electronic properties of graphene},
  author = {Castro Neto, A. H. and Guinea, F. and Peres, N. M. R. and Novoselov, K. S. and Geim, A. K.},
  journal = {Rev. Mod. Phys.},
  volume = {81},
  issue = {1},
  pages = {109--162},
  numpages = {0},
  year = {2009},
  month = {Jan},
  publisher = {American Physical Society},
  doi = {10.1103/RevModPhys.81.109},
  url = {https://link.aps.org/doi/10.1103/RevModPhys.81.109}
}

@article{neufeld2019floquet,
  title={Floquet group theory and its application to selection rules in harmonic generation},
  author={Neufeld, Ofer and Podolsky, Daniel and Cohen, Oren},
  journal={Nature Communications},
  volume={10},
  number={1},
  pages={405},
  year={2019},
  publisher={Nature Publishing Group UK London}
}

@article{batista2026field,
  title={Field-driven helicity in solid-state high-harmonic generation},
  author={Batista, Carlos and Menotti, Jean Paul and Kim, Dasol and Das, Bikash Kumar and Gao, Wenlong and Chac{\'o}n, Alexis and Granados, Camilo},
  journal={arXiv preprint arXiv:2604.27750},
  year={2026}
}

@article{reich2016rotating,
  title={Rotating-frame perspective on high-order-harmonic generation of circularly polarized light},
  author={Reich, Daniel M and Madsen, Lars Bojer},
  journal={Physical Review A},
  volume={93},
  number={4},
  pages={043411},
  year={2016},
  publisher={APS}
}

@article{reich2016illuminating,
  title={Illuminating molecular symmetries with bicircular high-order-harmonic generation},
  author={Reich, Daniel M and Madsen, Lars Bojer},
  journal={Physical Review Letters},
  volume={117},
  number={13},
  pages={133902},
  year={2016},
  publisher={APS}
}

@article{neufeld2021light,
  title={Light-driven extremely nonlinear bulk photogalvanic currents},
  author={Neufeld, Ofer and Tancogne-Dejean, Nicolas and De Giovannini, Umberto and H{\"u}bener, Hannes and Rubio, Angel},
  journal={Physical Review Letters},
  volume={127},
  number={12},
  pages={126601},
  year={2021},
  publisher={APS}
}

@article{lesko2026probing,
  title={Probing Broken Time-Reversal Symmetry in 2D Materials with Tailored-Light Photocurrent Generation},
  author={Lesko, Daniel MB and Weitz, Tobias and Wittigschlager, Simon and Nocker, Selina and Li, Weizhe and Hommelhoff, Peter and Neufeld, Ofer},
  journal={ACS nano},
  volume={20},
  number={15},
  pages={11614--11623},
  year={2026},
  publisher={ACS Publications}
}

@article{Keldysh:1965ojf,
    author = "Keldysh, L. V.",
    title = "{Ionization in the Field of a Strong Electromagnetic Wave}",
    journal = "Sov. Phys. JETP",
    volume = "20",
    pages = "1307--1314",
    year = "1965"
}

@article{molinero2025semiconductor,
  title={Semiconductor Wannier equations: a real-time, real-space approach to the nonlinear optical response in crystals (ATATA)},
  author={Molinero, Eduardo B and Amorim, Bruno and Ivanov, Misha and Brown, Graham G and Cistaro, Giovanni and Lopes, Jo{\~a}o M and Jim{\'e}nez-Gal{\'a}n, {\'A}lvaro and San-Jose, Pablo and Silva, Rui EF},
  journal={arXiv preprint arXiv:2510.22064},
  year={2025}
}

@article{cayssol2013floquet,
  title={Floquet topological insulators},
  author={Cayssol, J{\'e}r{\^o}me and D{\'o}ra, Bal{\'a}zs and Simon, Ferenc and Moessner, Roderich},
  journal={Physica Status Solidi (RRL)--Rapid Research Letters},
  volume={7},
  number={1-2},
  pages={101--108},
  year={2013},
  publisher={Wiley Online Library}
}

@article{rudner2020band,
  title={Band structure engineering and non-equilibrium dynamics in Floquet topological insulators},
  author={Rudner, Mark S and Lindner, Netanel H},
  journal={Nature Reviews Physics},
  volume={2},
  number={5},
  pages={229--244},
  year={2020},
  publisher={Nature Publishing Group UK London}
}

@article{haldane1988model,
  title = {Model for a Quantum Hall Effect without Landau Levels: Condensed-Matter Realization of the "Parity Anomaly"},
  author = {Haldane, F. D. M.},
  journal = {Phys. Rev. Lett.},
  volume = {61},
  issue = {18},
  pages = {2015--2018},
  numpages = {0},
  year = {1988},
  month = {Oct},
  publisher = {American Physical Society},
  doi = {10.1103/PhysRevLett.61.2015},
  url = {https://link.aps.org/doi/10.1103/PhysRevLett.61.2015}
}

@article{zeng2012valley,
  title={Valley polarization in MoS2 monolayers by optical pumping},
  author={Zeng, Hualing and Dai, Junfeng and Yao, Wang and Xiao, Di and Cui, Xiaodong},
  journal={Nature Nanotechnology},
  volume={7},
  number={8},
  pages={490--493},
  year={2012},
  publisher={Nature Publishing Group UK London}
}

@article{rana2023all,
  title={All-optical ultrafast valley switching in two-dimensional materials},
  author={Rana, Navdeep and Dixit, Gopal},
  journal={Physical Review Applied},
  volume={19},
  number={3},
  pages={034056},
  year={2023},
  publisher={APS}
}

@article{ang2017valleytronics,
  title={Valleytronics in merging Dirac cones: All-electric-controlled valley filter, valve, and universal reversible logic gate},
  author={Ang, Yee Sin and Yang, Shengyuan A and Ma, Zhongshui},
  journal={Physical Review B},
  volume={96},
  number={24},
  pages={245410},
  year={2017},
  publisher={APS}
}

@article{gomez2013floquet,
  title = {Floquet-Bloch Theory and Topology in Periodically Driven Lattices},
  author = {G\'omez-Le\'on, A. and Platero, G.},
  journal = {Phys. Rev. Lett.},
  volume = {110},
  issue = {20},
  pages = {200403},
  numpages = {5},
  year = {2013},
  month = {May},
  publisher = {American Physical Society},
  doi = {10.1103/PhysRevLett.110.200403},
  url = {https://link.aps.org/doi/10.1103/PhysRevLett.110.200403}
}

@article{heide2024petahertz,
  title={Petahertz electronics},
  author={Heide, Christian and Keathley, Phillip D and Kling, Matthias F},
  journal={Nature Reviews Physics},
  volume={6},
  number={11},
  pages={648--662},
  year={2024},
  publisher={Nature Publishing Group UK London}
}

@article{ghimire2019high,
  title={High-harmonic generation from solids},
  author={Ghimire, Shambhu and Reis, David A},
  journal={Nature physics},
  volume={15},
  number={1},
  pages={10--16},
  year={2019},
  publisher={Nature Publishing Group UK London}
}

@article{Ciappina_2017,
	Author = {Ciappina, M F and P{\'e}rez-Hern{\'a}ndez, J A and Landsman, A S and Okell, W A and Zherebtsov, S and F{\"o}rg, B and Sch{\"o}tz, J and Seiffert, L and Fennel, T and Shaaran, T and Zimmermann, T and Chac{\'o}n, A and Guichard, R and Za{\"\i}r, A and Tisch, J W G and Marangos, J P and Witting, T and Braun, A and Maier, S A and Roso, L and Kr{\"u}ger, M and Hommelhoff, P and Kling, M F and Krausz, F and Lewenstein, M},
	Doi = {10.1088/1361-6633/aa574e},
	Journal = {Reports on Progress in Physics},
	Month = {mar},
	Number = {5},
	Pages = {054401},
	Publisher = {IOP Publishing},
	Title = {Attosecond physics at the nanoscale},
	Url = {https://doi.org/10.1088/1361-6633/aa574e},
	Volume = {80},
	Year = {2017}}

@article{PhysRevLett.113.073901,
  title = {Theoretical Analysis of High-Harmonic Generation in Solids},
  author = {Vampa, G. and McDonald, C. R. and Orlando, G. and Klug, D. D. and Corkum, P. B. and Brabec, T.},
  journal = {Phys. Rev. Lett.},
  volume = {113},
  issue = {7},
  pages = {073901},
  numpages = {5},
  year = {2014},
  month = {Aug},
  publisher = {American Physical Society},
  doi = {10.1103/PhysRevLett.113.073901},
  url = {https://link.aps.org/doi/10.1103/PhysRevLett.113.073901}
}

@article{PhysRevB.111.155421,
  title = {Valley filtering and valley valves in irradiated pristine graphene},
  author = {Kumari, Rekha and Dixit, Gopal and Kundu, Arijit},
  journal = {Phys. Rev. B},
  volume = {111},
  issue = {15},
  pages = {155421},
  numpages = {13},
  year = {2025},
  month = {Apr},
  publisher = {American Physical Society},
  doi = {10.1103/PhysRevB.111.155421},
  url = {https://link.aps.org/doi/10.1103/PhysRevB.111.155421}
}

\end{document}